# Machine Learning for Uncovering Biological Insights in Spatial Transcriptomics Data

Alex J. Lee[1*], Robert Cahill[1*], Reza Abbasi-Asl[1#]

[1]University of California, San Francisco

*These authors contributed equally to this work.
#Corresponding author: Reza Abbasi-Asl (Email: Reza.AbbasiAsl@ucsf.edu)

## Abstract

Development and homeostasis in multicellular systems both require exquisite control over spatial molecular pattern formation and maintenance. Advances in spatially-resolved and high-throughput molecular imaging methods such as multiplexed immunofluorescence and spatial transcriptomics (ST) provide exciting new opportunities to augment our fundamental understanding of these processes in health and disease. The large and complex datasets resulting from these techniques, particularly ST, have led to rapid development of innovative machine learning (ML) tools primarily based on deep learning techniques. These ML tools are now increasingly featured in integrated experimental and computational workflows to disentangle signals from noise in complex biological systems. However, it can be difficult to understand and balance the different implicit assumptions and methodologies of a rapidly expanding toolbox of analytical tools in ST. To address this, we summarize major ST analysis goals that ML can help address and current analysis trends. We also describe four major data science concepts and related heuristics that can help guide practitioners in their choices of the right tools for the right biological questions.

## Parallel advances in spatial transcriptomics and machine learning

The parallel and rapid technology advances in both spatial transcriptomics (ST) and machine learning (ML) present an unprecedented opportunity to enhance our understanding of biology of cells, tissues, development and disease. Researchers can now identify spatial distributions of specific cell-types and the variability of specific gene of interest with in-situ hybridization techniques such as MERFISH[1] and smFISH[2,3]. Next-generation sequencing based techniques such as Visium[4], GeoMx[5], and XYZeq[6] allow unbiased spatial characterization of the whole transcriptome. Technology development in ST is rapid, and advances are expected to continue across three dimensions: higher resolution, integration with other data modalities, and robust quantification of spatiotemporal dynamics[7,8]. Rapid advances in ML are allowing biological researchers to capitalize on the increasingly large and complex ST datasets to facilitate biological discovery. Recently, ML's deep learning subfield has opened new doors in biology, most notably AlphaFold's dramatic impact on previously intractable problems in structural biology[9–12]. Methods leveraging convolutional, graph, and transformer-based neural networks are now amongst the state-of-the-art in areas as diverse as histology, multiplexed imaging, and gene and protein interaction network analysis [13–17].

Here, we offer a practical discussion of opportunities, trade-offs, and pitfalls in the ML-based analysis of spatial transcriptomics data and integration with other datasets. In order to facilitate clear descriptions of the trade-offs, we identify four relevant concepts in biological data science (accuracy, interpretability, stability, and computability)[18,19], and apply these concepts to ML for ST. Here, we aim to describe the ST questions enabled by ML, the key ML techniques and toolboxes for ST, and guidance for developing and applying these techniques based on the four principles of biological data science.

## Machine learning as a fundamental tool for spatial transcriptomics data analysis

There has been a profusion of new ML tools for ST, with dozens of ML tools now publicly available and new methods published on a monthly basis. We describe three categories of biological questions in ST data that can be investigated more effectively using ML and note a broad, but nonexhaustive list of ML toolboxes in each group (**Fig. 1**).

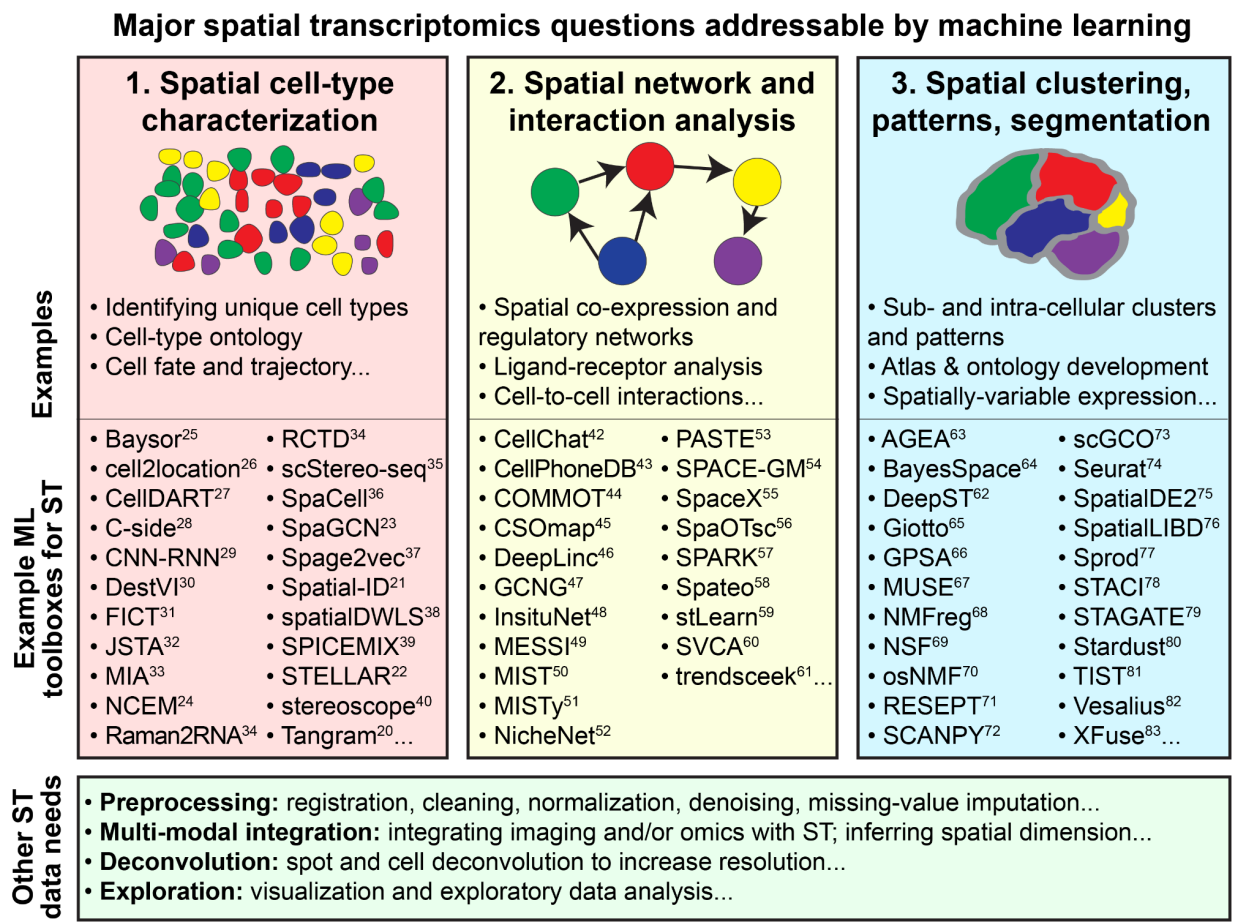

**Fig 1. Critical questions in spatial transcriptomics (ST) that are addressable by machine learning (ML).** ML allows the exploration of ST data to address fundamental questions, including 1) spatial cell-type characterization[20–41], 2) spatial network, regulatory, and interaction analysis[42–61], and 3) spatial clustering, patterns, and segmentation[62–83]. The specific ML toolboxes are numerous, and ever-growing. Many toolboxes are designed for one specific ST-related question, and others can address all three ST questions.

**ML for spatial cell-type characterization:** First, how can we utilize spatial information to enrich our understanding of cell-type diversity, and then further leverage this information to understand higher-order structure in tissues and organs? Spatial cell type characterization facilitates understanding of the heterogeneity, development, and interaction of tissues and biological systems. ML solutions will be crucial in systematically dissecting the impact of spatial versus non-spatial variability in the development and maintenance of cellular identity. A broad class of methods in this area performs cell-type transfer from annotated scRNA-seq datasets. This can be done via learning projection or similarity functions that map from single-cell reference profiles onto spatial domains[20]. For example, given a reference scRNA-seq cell gene expression profile, a method may compute a probability of a reference cell type occurring in a puck or directly compute the probability of a query cell being of that reference cell type by computing a similarity measurement. Graph neural networks or operations on spatial graphs of cells are often utilized to allow explicitly for spatial information to inform this mapping[21,22]. These methods are often enabled by supervised operations on cellular neighborhood graphs to optimize clustering[23] or other properties such as cell-cell similarity with a reference. Although these methods frequently require supervision in the form of single-cell reference profiles, they can also incorporate an unsupervised low-dimensional representation learning step using principal component analysis (PCA) or using methods such as variations autoencoders (VAEs). A newer approach for low-dimensional cellular representation learning utilizes self-supervised learning, where a cell neighborhood graph is constructed and a graph neural network is used to predict a cell's gene expression profile from neighboring cells[22,24].

**ML for spatial network, regulatory, and interaction analysis:** Second, what are the spatial interactions and networks that underlie cell-type diversity? ST and related techniques offer unprecedented capabilities to analyze and understand the impact of cellular proximity on communication at scale. Similar to spatial cell-type annotation, methods for cell interaction analysis often require supervision in the form of pre-definition of ligand-receptor pairs by a user or from a database. Once these pairs are defined, quantification of the interaction between specific ligand-receptor pairs or groups can be formulated as a simpler statistical association or prediction problem. Common frameworks include optimal transport and graph affinity-based algorithms. Optimal transport methods compute a notion of dissimilarity between probability distributions of ligands or receptors[44]. Graph affinity algorithms represent ligands and receptors as nodes in a graph and similarity can be computed by affinity metrics such as likelihood of a random walk proceeding from a ligand node to a receptor node[52]. Often, interaction analysis is posed in terms of identifying two clusters of cells associated with statistically unlikely expression of ligand-receptor pairs in the two clusters[43,59]. Graph neural networks are a common tool for spatial cell interaction analysis both for generating low-dimensional cell representations for clustering and for direct prediction of cell interactions[46].

**ML for spatial clustering, segmentation, and patterns recognition:** Third, how can we identify patterns, clusters, and segments of tissues and biological regions? ML can be used to create flexible yet powerful models that can incorporate expert domain-knowledge about the statistical structure of these different regions in tissue. Frequently used methods include statistical tools such as a Gaussian process to identify a low number of latent spatial factors that give rise to spatial correlations between cells, spots, or pucks[69,75]. A Gaussian process approach can also be used to identify spatially varying genes by examining the relative contribution of the latent factors to each gene's observed spatial pattern. Statistical models such as Leiden clustering[72,74] or hidden Markov random fields[65] are also commonly used. Another approach is to learn a low dimensional representation of cells using a neural network or unsupervised dimensionality reduction technique with or without integration of spatial information and perform clustering. Graph and convolutional neural networks are commonly employed, and can be used to implement assumptions; for example, by constraining the distance at which cells are modeled as neighbors in neighborhood graphs.

**Other ML-based data processing needs for ST:** In addition to the three major biological questions, we highlight other data processing needs for ST, which are also common for non-spatial transcriptomics and other data science pipelines (**Fig. 1**). These data preprocessing applications include data curation, registration, normalization, denoising, missing-value imputation, and others[84–88]. The addition of a data-driven ML component to these workflows has enabled more robust, flexible, and accurate preprocessing in the field of genomics and consequently ST. This is in part due to the adaptability of ML-based preprocessing pipelines to the growing amount of data available to train them, as well as the community's creativity in posing preprocessing steps as ML-compatible workflows such as prediction or latent variable discovery problems.

To summarize, there has been a profusion of ML toolboxes for ST in the last few years to address core biological questions and help with various data processing needs.

## Practical considerations for machine learning-enabled discoveries in spatial transcriptomics

Building on the recent developments in the field of interpretable machine learning[18,19], we describe four relevant concepts in data science that are critical in successful ML-enabled discoveries for ST data: predictive accuracy, interpretability, stability, and computability (**Fig. 2**). Although accuracy is frequently discussed in methodological papers, discussions of interpretability, stability, and computability are relatively less frequent. The first concept, predictive accuracy, describes the ML model's ability to deliver predictions or classifications that reflect experimental measurements and ideally biological reality[18]. The second concept, model interpretability, is critically important to address the challenge of "black box" or "gray box" ML models, where the difficulty in explaining why an ML model makes a certain prediction limits its usefulness for discovery in the biological sciences[89]. While interpretability has widely varying meanings, it generally refers to the ability to extract domain-relevant knowledge that is contained in the data or learned by the model[19]. Some of the recent ML tools in ST incorporate interpretability in their model and method design[39,51,70,71], however, interpretability has yet to become a central criteria in scientific oriented ML modeling. The third concept is stability, which is a measure of scientific and statistical reproducibility; it asks whether each step of the ML pipeline produces consistent results with slight perturbations in the model or data[18,19]. Stability has been established as a minimum requirement for interpretability[90,91,70]. Increased consideration of stability analysis in ML may facilitate increased reproducibility and interpretability. Fourth, computability is a "gatekeeper of data science", and refers here to the computational feasibility of ML models and algorithms[18]. Computability is often described in terms of algorithmic efficiency, time/space complexity, and overall runtime. ST data resolution and overall data size and complexity are growing rapidly[92]. Further, modern ML tools based on deep learning techniques are often orders of magnitude more computationally complex than classic ML tools. Together, this creates computability challenges, especially for labs without deep expertise in high-performance computing and parallel processing. It also provides an opportunity for collaboration across disciplines. Team science may be a solution, where computational-focused labs partner with biological-focused labs to set up high-performance pipelines. Several of the ML toolboxes described in **Fig. 1** incorporate specific design constraints for computational speedup: examples include sparsity matrix algorithm design[21] or adopting novel heuristics for model fitting[73].

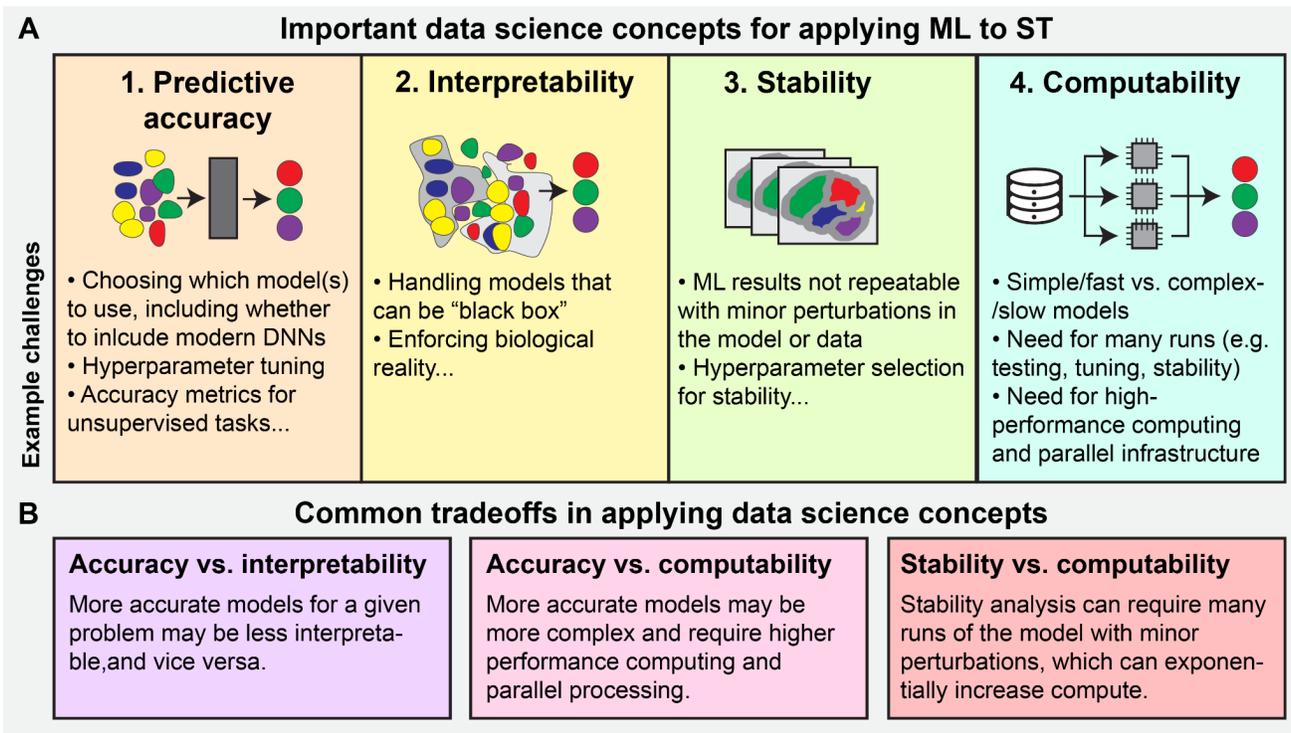

**Figure 2. Principles and approaches for applying ML to ST data.** Predictive accuracy, interpretability, stability, and computability are important principles for biological data science. Adhering to these principles will take thoughtful planning and tradeoffs in model choice and implementation.

These four concepts are critical, yet sometimes are in opposition to each other, requiring tradeoffs to be made (**Fig. 2B**). However, a careful consideration of the four critical concepts and the goals of the project in question can inform the choices of the employed method. One example is the trade-off between accuracy and interpretability. For example, if the goal of an analysis is to perform cell-type transfer from scRNA-seq data to those observed in ST, our desire for high accuracy may outweigh our desire for model interpretability. However, for analyses that will be used to facilitate human understanding of the fundamental behavior of a biological system, it may be more important to incorporate explainability and interpretability. For example, many methods in spatial region finding allow the user to directly visually inspect factors or putative spatially important patterns[69,70]. When coupled with methods that allow inspection of the statistical or machine learning model that generated these spatial patterns, researchers can develop a reasonable understanding of the patterns that our models learn, facilitating understanding of the biological system and potentially allowing improved downstream analysis. Conversely, our desire for predictive accuracy can also reduce the importance of stability, as long as the instability of our predictions is below some threshold (or the absolute performance is low), these factors may be less important.

Another aspect in which the goals of an ML-based analysis inform the choice of methods is in exploratory versus hypothesis driven research questions. Several methods in cell communication incorporate prior knowledge by directly letting known biology inform the method (for example, if a method may use an existing ligand-receptor pair database or an annotation transfer method uses specific marker genes that are user-provided to drive annotation). However, these methods can often implicitly preclude discovery of novel ligand-receptors or cell-types in observed data.

In some applications, it might be preferable to use a highly predictive but less interpretable method if the overall goal of the project is to integrate an exploratory analysis phase with an experimental validation stage. If an analysis goal is to identify novel cell types and marker genes from a perturbation-based experiment, it may be less important to have an easily interpretable method than to have in general a method with a low false negative rate, because any putative relationships are meant for immediate experimental testing.

Finally, the way in which a method will be used can mitigate its drawbacks, particularly if computability is deprioritized. As a simple example, least absolute shrinkage and selection operator (LASSO) regression is a commonly adopted tool for feature subset selection and analysis. LASSO uses a regularization parameter on the traditional least-squares regression model to shrink parameter estimates for all but a few covariates to zero. LASSO is unstable in that at a given regularization strength, the model will select one of several correlated variables at random. However, by producing an ensemble of models one may still use LASSO for feature selection for interpretable prediction[93]. At the expense of computational complexity, one can take advantage of this model stability for feature selection or hyperparameter selection for any of the models described in the previous section (for example, the number of clusters in a spatial clustering workflow). An exploration and understanding of the behavior of ensembles of models can be highly useful for exploring the stability versus accuracy tradeoffs of a given model. Variability in performance may not be apparent without this sort of analysis, and even if a model is relatively performant, if stability is low then that method may be less suitable for biological discovery. For example, if a machine learning method for spatially important gene selection produces highly varying selections of genes, it may be difficult to interpret whether these genes are biologically relevant in a given tissue instead of statistical outliers.

These tradeoffs may get more extreme as data sets grow and the new deep neural network architectures requiring larger datasets become mainstream. Therefore, it is crucial that the next-generation of ML models for spatial transcriptomics data takes into account and effectively addresses these tradeoffs. These considerations can ensure that ML-based models for ST data are not only accurate but also potentially explainable, stable, and computable.

## Conclusion

In this perspective, we provide an overview of opportunities and challenges of ML-enabled discoveries for ST data. ML is already a critical tool to analyze ST data and to address important biological questions around cell type, interactions, and regional segmentation. Methods based on modern ML techniques such as deep learning are starting to make their mark in ST. However, there are challenges with ML related to ST data, including around the data science concepts of accuracy, interpretability, stability, and computability. Practical consideration of these concepts along with their tradeoffs in ML model choice, implementation, and the development could enable a more effective integration of ML with scientific discovery.

## Acknowledgments

RA would like to acknowledge support from the National Institute of Mental Health of the National Institutes of Health under award number RF1MH128672, Weill Neurohub, and Sandler Program for Breakthrough Biomedical Research, which is partially funded by the Sandler Foundation.

## Competing Interests

The authors declare no competing interests.